\documentclass[apj]{emulateapj}

\usepackage{natbib}

\usepackage{graphicx}
\usepackage[percent]{overpic}
\usepackage{lipsum}
\usepackage{multirow}
\usepackage{amssymb}
\usepackage{color}
\usepackage{rotating}
\usepackage{mwe,tikz}
\usepackage[percent]{overpic}
\usepackage{mathtools}
\usepackage{physics}
\usepackage{amsmath}
\usepackage[utf8x]{inputenc}
\usepackage{hyperref}
\usepackage{wasysym}
\usepackage{soul}

\newcommand{\msun}{${\rm M_{\sun}}$}
\def\ltsima{$\; \buildrel < \over \sim \;$}
\def\simlt{\lower.5ex\hbox{\ltsima}}
\def\gtsima{$\; \buildrel > \over \sim \;$}
\def\simgt{\lower.5ex\hbox{\gtsima}}
\def\km{{\rm\,km}}
\def\kms{{\rm\,km\,s^{-1}}}
\def\km2s2{{\rm\,km^{2}\,s^{-2}}}
\def\pc{{\rm\,pc}}
\def\kpc{{\rm\,kpc}}

\def\msun{{\rm\,M_\odot}}


\def\deg{^\circ}

\def\Gyr{{\rm\,Gyr}}

\def\ltsima{$\; \buildrel < \over \sim \;$}
\def\gtsima{$\; \buildrel > \over \sim \;$}

\def\cocoon{{\it\,cocoon\,}}

\slugcomment{Accepted by The Astrophysical Journal Letters}

\shorttitle{Phase-space correlation in cold stellar streams}
\shortauthors{Malhan et al.}

\begin{document}

\title{Phase-space correlation in stellar streams of the Milky Way halo:\\ The clash of Kshir and GD-1\altaffilmark{1}}

\email{khyati.malhan@fysik.su.se}

\author{Khyati Malhan\altaffilmark{2}, Rodrigo A. Ibata\altaffilmark{3}, Raymond G. Carlberg\altaffilmark{4}, Michele Bellazzini\altaffilmark{5}, Benoit Famaey\altaffilmark{3} and Nicolas F. Martin\altaffilmark{3}}

\altaffiltext{1}{Based on observations obtained at the Canada-France-Hawaii Telescope (CFHT) which is operated by the National Research Council of Canada, the Institut National des Sciences de l´Univers of the Centre National de la Recherche Scientique of France, and the University of Hawaii.}

\altaffiltext{2}{The  Oskar  Klein  Centre  for  Cosmoparticle  Physics,  Department  of Physics,  Stockholm  University,  AlbaNova,  10691  Stockholm,  Sweden}
\altaffiltext{3}{Universit\'e de Strasbourg, CNRS, Observatoire Astronomique de Strasbourg, UMR 7550, F-67000 Strasbourg, France}
\altaffiltext{4}{Department of Astronomy \& Astrophysics, University of Toronto, Toronto, ON M5S 3H4, Canada}
\altaffiltext{5}{INAF - Osservatorio di Astrofisica e Scienza dello Spazio, via Gobetti 93/3, 40129 Bologna, Italy}

\begin{abstract}
We report the discovery of a $70\deg$ long stellar stream in the Milky Way halo, which criss-crosses the well known ``GD-1'' stream.  We show that this new stellar structure (``Kshir'') and GD-1 lie at similar distance, and are remarkably correlated in kinematics. We propose several explanations for the nature of this new structure and its possible association with GD-1. However, a scenario in which these two streams were accreted onto the Milky Way within the same dark matter sub-halo seems to provide a natural explanation for their phase-space entanglement, and other complexities of this coupled-system.
\end{abstract}

\section{Introduction}\label{sec:Introduction}

Stellar streams are ``fossil'' remnants of accretion events that are formed by the tidal disruption of satellite systems as they accrete and begin to orbit in the gravitational potential of the host galaxy. More than 50 streams have been detected so far in the Milky Way halo (for e.g., \citealt{Ibata2001Sgr, Belokurov2006, Grillmair2009_fourStreams, Bernard2016, Balbinot2016Phoenix, Myeong2017_Streams, Shipp2018, Malhan_Ghostly_2018, Ibata_Norse_streams2019}). A large fraction of these old and metal poor structures are observed as narrow and one-dimensional structures \citep{GrillmairCarlin2016}, and are explained as stellar debris produced from globular clusters (GCs), that were perhaps brought in by their parent dark satellite galaxies during accretion \citep{Renaud2017}. Due mainly to their simple morphology and lack of associations with any other observed components of the stellar halo, the GC streams are generally modeled independently as simple GCs disrupting under the tidal force field of the host galaxy (e.g., \citealt{Dehnen2004}). The good match between observations and such simple models, in effect, also favours primeval models of GC formation - a scenario in which GCs originate from dark matter free gravitationally-bound gas clouds in the early Universe \citep{Kravtsov2005, Kruijssen2014}, and then later migrate into the host galaxy. 

In this work, we report the discovery of a new Milky Way stream, and find that it criss-crosses through the previously well known ``GD-1'' stream. We demonstrate that GD-1 and this new structure are highly correlated in distance and kinematics, and have similar stellar populations. 
We discuss the possible interpretations of the origin of this remarkable entangled system, and the potential implications of our results in dark matter and GC formation studies.

\begin{figure*}
\begin{center}
\vspace{-0.3cm}
\includegraphics[width=0.94\hsize]{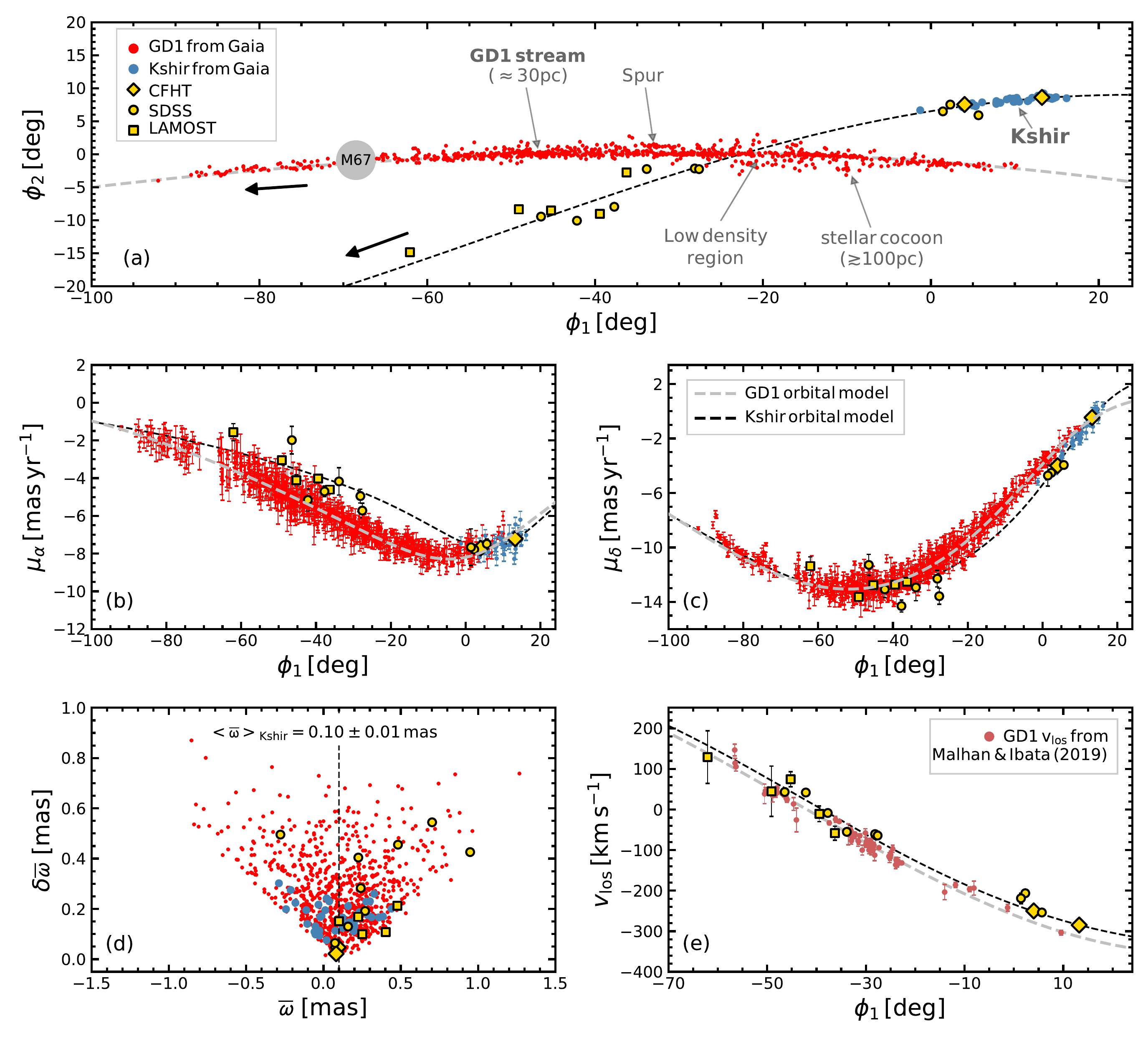}
\end{center}
\vspace{-0.5cm}
\caption{Spatial and kinematic distribution of GD-1 and Kshir. a: Sky position in $\phi_1-\phi_2$ coordinates, which are a rotated celestial system aligned along GD-1. The narrow GD-1 ($\approx 30$ pc wide) can be immediately spotted along $\phi_2 \sim 0\deg$. Some additional features can also be observed, such as the low density regions along the stream, the ``spur'' component and the ``cocoon'' component ($\simgt 100$ pc wide). An arc-like structure can be seen at $(\phi_1,\phi_2) \sim (10\deg,8\deg)$, which we refer to as ``Kshir''. Spectroscopically confirmed members for Kshir are shown in yellow. The region of  sky containing the foreground open cluster M67 $(d_{\odot}\sim 0.9\kpc)$ was masked out prior to the running of the \texttt{STREAMFINDER}. The bold arrows indicate the direction of motion of the two streams. Panels b, c and d show, respectively, proper motion in $\mu_{\alpha}$, $\mu_{\delta}$, and parallax $\overline{\mathbb{\omega}}$, as a function of $\phi_1$. Panel e shows the heliocentric line-of-sight velocities $v_{\rm los}$ of the members of GD-1 (pink) and Kshir (yellow). The derived orbits of the two structures are shown in each panel, and can be seen to be very similar.}
\label{fig:Fig_1_new}
\end{figure*}
\section{GD-1 and its neighbour}\label{sec:GD1_and_sibling}

Ranging in heliocentric distance between $d_{\odot} \sim 8-12\kpc$, the GD-1 stream \citep{GrillmairGD12006} has been observed as an $\sim 80\deg$ long ($\sim 12\kpc$, \citealt{WhelanBonacaGD12018}), narrow ($\approx 20\pc$ in physical width, \citealt{Koposov2010}), linear stellar structure. This GC stream is dynamically very cold (with a velocity dispersion of $\approx 1 \kms$, \citealt{Malhan2018PotentialGD1}, MI19 hereafter) and is remarkably metal deficient ([Fe/H]$= -2.24\pm0.21$ dex).

Fig~\ref{fig:Fig_1_new}a shows the density map in the region around GD-1 that we obtained by processing the entire ESA/Gaia DR2 datatset \citep{GaiaDR2_2018_Brown, GaiaDR2_2018_astrometry} with the \texttt{STREAMFINDER} software \citep{Malhan2018_SF, Malhan_Ghostly_2018, Ibata_Norse_streams2019}. Briefly, \texttt{STREAMFINDER} works by examining every star in the Gaia survey in turn, sampling the possible orbits consistent with the observed photometry and kinematics, and finding the maximum-likelihood stream solution given a contamination model and a stream model. As in \cite{Malhan_Ghostly_2018}, we de-reddened the survey using the \cite{Schlegel1998} dust maps, and kept only those stars with (de-reddened) $G_0<19.5$ to ensure homogeneous depth over the sky. The stream width parameter in the search algorithm was set to $50\pc$, and we used single stellar population (SSP) template models from the PARSEC stellar tracks library \citep{Parsec_isochrones2012} of age $12.5\Gyr$, and scanned a range of metallicities of ${\rm [Fe/H]}=-2.2, -2.0, -1.6, -1.2, -0.8, -0.4$. For each processed star, we thus obtained 6 stream solutions, corresponding to the 6 trial SSP metallicity values, and we accepted the solution that yielded the highest likelihood. GD-1 appears as a completely distinguished structure in our \texttt{STREAMFINDER} maps, as shown in Fig~\ref{fig:Fig_1_new}a. It transpires that the best orbital solution for 88\% of the GD-1 stars is obtained with an SSP template with metallicity ${\rm [Fe/H]}=-2.2$, similar to the measured [Fe/H] value (MI19)\footnote{The value 88\% corresponds to the fraction of stars (identified as GD-1) obtained as stream solution using that particular [Fe/H] model.}. We also set the stream-detection significance to $>8\sigma$, which means that at the position of every star, the algorithm finds that there is a $>8\sigma$ significance for there to be a stream-like structure. This results in a sample of $811$ stars in the region around GD-1 that are shown in Fig~\ref{fig:Fig_1_new}a (red and blue points). The region of the sky containing the foreground open cluster M67 was masked out prior to running the \texttt{STREAMFINDER}.

Fig~\ref{fig:Fig_1_new}a reveals the complex structure that surrounds the thin GD-1 stream (shown with red points). The plot is presented in $\phi_1 - \phi_2$ coordinates \citep{Koposov2010}, which align along GD-1. A narrow component of GD-1 ($\approx 30\pc$ wide) can be readily seen along $\phi_2 \approx 0\deg$, enveloped by a broad and diffuse structure ($\simgt 100\pc$ wide). This extended component was previously reported in \cite{MalhanCocoonDetection2019}, and was referred to as the \cocoon component. The code also tentatively detected the previously-known low density regions and the ``spur'' component \citep{WhelanBonacaGD12018}. 

The present study focuses on the detection of the arc-like feature that is conspicuously visible in Fig~\ref{fig:Fig_1_new}a at $(\phi_1,\phi_2) \approx (10\deg,8\deg)$, skirting almost parallel to GD-1 (shown with blue points). We refer to this structure as ``Kshir''\footnote{In Hindu mythology, {\it Kshir} refers to a water body made of milk.}. Based on the orbital analysis of the \texttt{STREAMFINDER} code, we note that both Kshir and GD-1 are found by the algorithm with similar values of the $z$-component of angular momentum ($L_z$) and energy ($E$). In particular, the algorithm (in the adopted potential model) estimates $(L_z, E)_{\rm GD1}=(2900\pm800\kpc\kms, -88000\pm19000\km2s2)$ and $(L_z, E)_{\rm Kshir}=(3200\pm600\kpc\kms, -89000\pm13000\km2s2)$. This indicates that they are, perhaps, part of the same coherent group and share a common origin. The \texttt{STREAMFINDER} detects a total of $42$ stars for the Kshir stream from Gaia DR2. We realized that one of these stars fortuitously had a spectroscopic observation in the SDSS/Segue (DR10, \citealt{SEGUE_SDSS2009}), from which we obtained the metallicity (${\rm [Fe/H]}$) and line-of-sight velocity $(v_{\rm los})$ values. A further 2 stars were observed with the ESPaDOnS high-resolution spectrograph on the Canada-France-Hawaii Telescope (CFHT) in service mode, as a part of our own follow up program. The data were reduced with the Libre-ESpRIT pipeline \citep{1997MNRAS.291..658D}, and we measured the stars' velocities using the {\tt fxcor} command in IRAF. The chemical abundances of the stars (which are part of a much larger sample of streams) are currently being analysed, and will be presented in a later contribution. These $3$ stars are mentioned in the top rows of Table~\ref{tab:GD1x_inventory} and are also shown in Fig~\ref{fig:Fig_1_new}. Kshir is already visible in Fig~1 of \cite{MalhanCocoonDetection2019}, however was not focussed upon, as we previously lacked spectroscopic measurements for this structure.

We used the Kshir orbit model, derived in Section~\ref{sec:orbits}, to find additional member stars that lie along its trajectory. To this end, we used the 5D astrometric measurements (that came from Gaia DR2 for blue stars shown in Fig~\ref{fig:Fig_1_new}a), in combination with the aforementioned $3$ velocity measurements, for Kshir stars to obtain a solution for its orbit. This orbit-fitting procedure is detailed in Section~\ref{sec:orbits}. We now assume that we possess a reasonable orbital representation of the Kshir structure.

\begin{figure}
\begin{center}
\vbox{
\includegraphics[width=0.90\hsize]{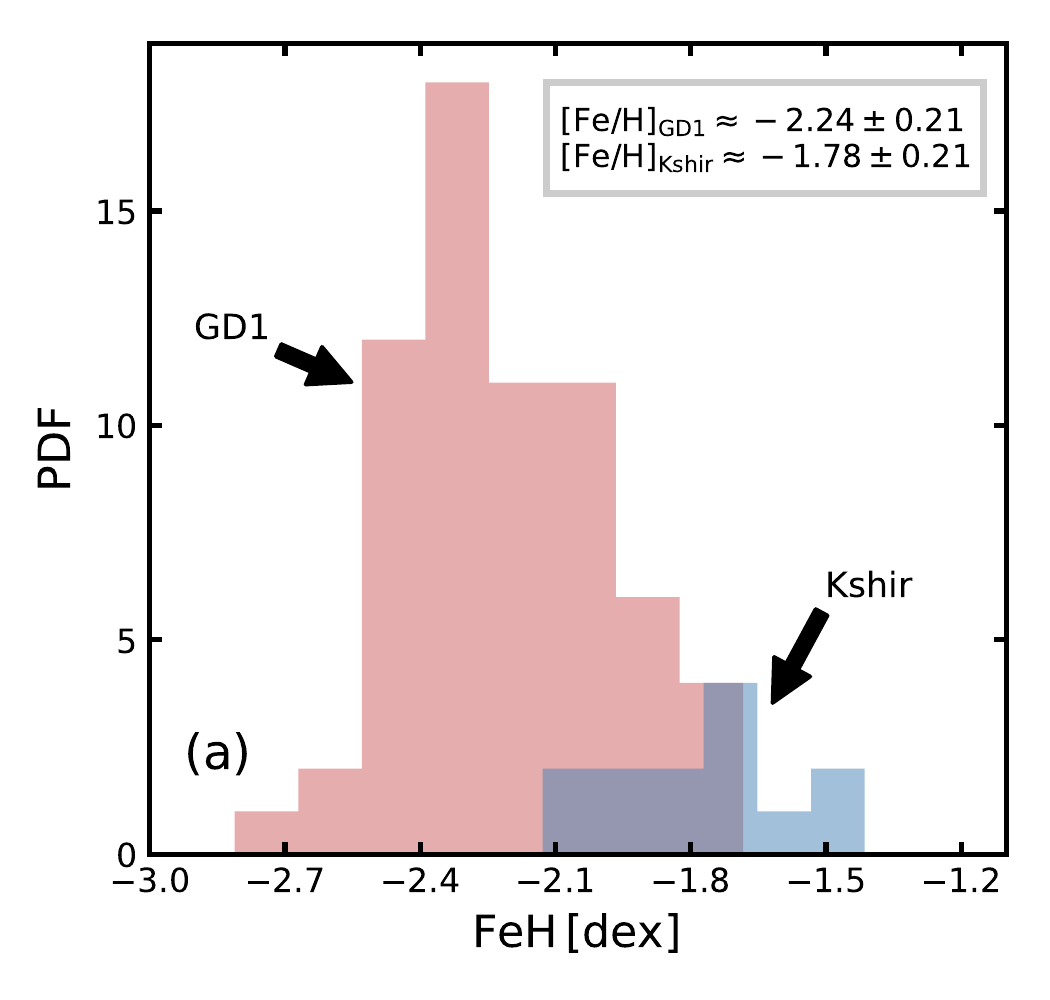}
\includegraphics[width=0.90\hsize]{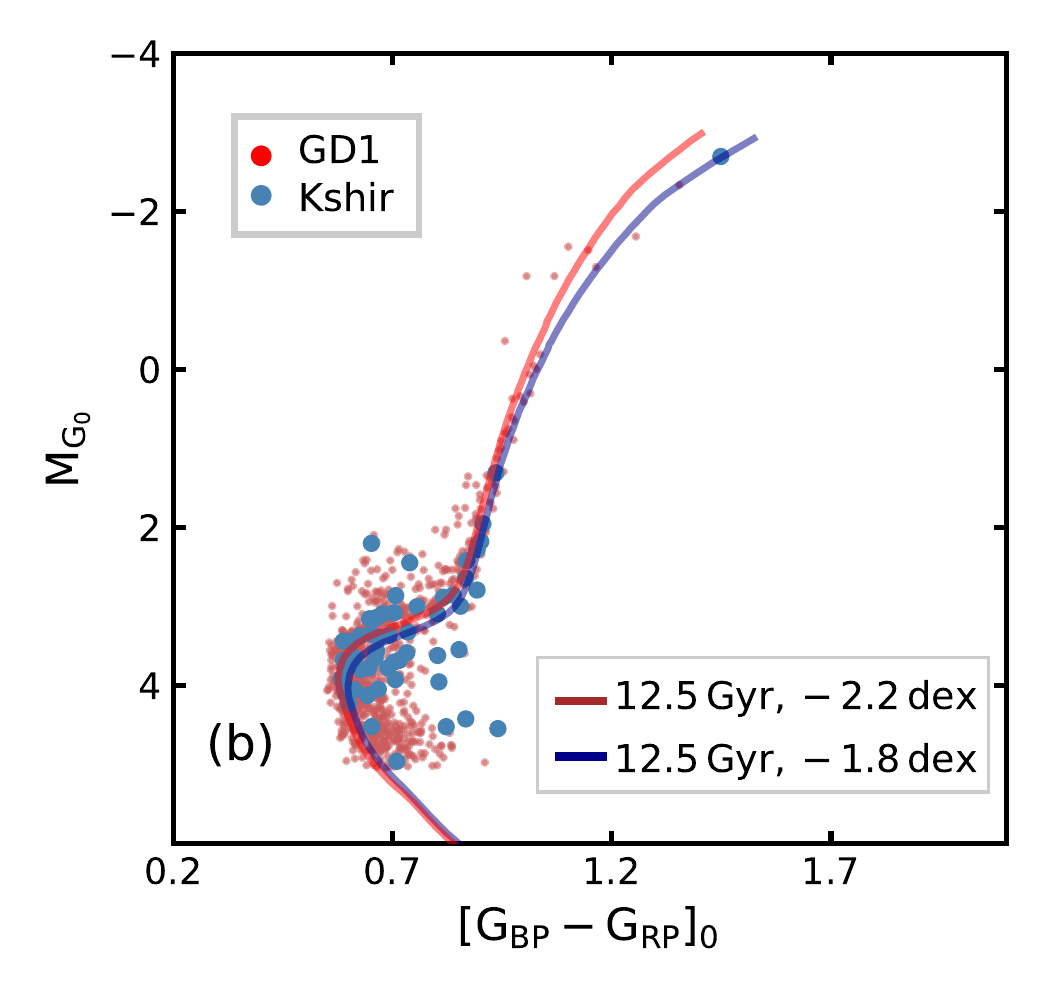}
}
\end{center}
\vspace{-0.5cm}
\caption{Metallicity and photometry of GD-1 and Kshir. {\it Top panel:} [Fe/H] distribution of spectroscopically-confirmed members of GD-1 (red) and Kshir (blue) stars. {\it Bottom panel:} Gaia (de-reddened) color-magnitude diagram of all member stars of Kshir and GD-1, previously shown in Fig~\ref{fig:Fig_1_new}. The absolute magnitude were then obtained by correcting the observed magnitude value for the orbital distance of each star. Two SSP template models, corresponding to the mean metallicity values of the streams, are also shown.}
\label{fig:Fig_chemistry}
\end{figure}

To identify additional spectroscopic members of Kshir, we first created a special dataset by cross-matching the Gaia DR2 catalogue with the SDSS \citep{SEGUE_SDSS2009} and LAMOST DR4 \citep{Lamost2012} spectroscopic datasets, and selected those stars with Gaia colors in the range $(0.40<[G_{\rm BP}−G_{\rm RP}]_0<1.60)$, with magnitude $(G_0 < 20)$, and with spectroscopic metallicity ${\rm [Fe/H]< -0.5\,dex}$. We further selected those stars that lie within $3\sigma$ of Kshir's orbit model in the observed parallax $(\overline{\mathbb{\omega}})$, proper motion  $(\mu_{\alpha}$, $\mu_{\delta})$, and line of sight velocity $v_{\rm los}$ space, and within $\sim 300 \pc$ perpendicular to the orbit (which defines the maximum possible stream width). We do not make any additional selections in photometry and [Fe/H] at this stage, since we do not a-priori know the properties of Kshir's stellar population. This 6-dimensional selection yields a total of $13$ stars ($8$ from SEGUE and $5$ from LAMOST), that represent additional Kshir's candidate members. These stars were not previously identified by the \texttt{STREAMFINDER} due to their lower contrast. We verified that the expected contamination from a smooth halo model (as predicted by the GUMS simulation, \citealt{GUMS2012_Robin}) along Kshir's orbit is nearly zero.

The combined phase-space properties of the stellar members of Kshir and GD-1 are plotted in Fig~\ref{fig:Fig_1_new}a-e (along with the spectroscopically confirmed members for Kshir that are shown in yellow). One readily observes that GD-1 and Kshir intersect spatially, at $\phi_1 \approx -20\deg$, and are also strongly entangled in proper motion space. In the $v_{\rm los}$ panel, the stars corresponding to GD-1 refer to the spectroscopically-confirmed members inventoried in MI19. It can be easily discerned that Kshir stars lie quite close to the GD-1 stars even in $v_{\rm los}$ space. The orbit of GD-1 (obtained from MI19) and Kshir look very similar in every phase-space dimension, and Kshir's orbit predicts similar $v_{\rm los}$ gradient along the length of the stream as observed for GD-1, with an almost constant offset of $\approx 20\kms$. We caution that the lack of stars between $\phi_1\sim-25\deg\,\rm{to}\,0\deg$ in Kshir may either be physical in origin, or could be due to the selection effect of the SEGUE and LAMOST surveys; however is hard to quantify at this stage. Also note that the leading part of Kshir (dominated by 6D members) appears much wider than the trailing part. This result could be specific to the criteria adopted here to select Kshir stars, and future analysis (with a larger sample size) should better characterise the structural morphology of this stream.

The uncertainty-weighted average mean parallax for Kshir (for blue points in Fig~\ref{fig:Fig_1_new}) is $<\overline{\mathbb{\omega}}>=0.10\pm 0.01$ mas, i.e. $\sim 10\kpc$ in distance, which is similar to the distance of the GD-1 orbit in the same range of $\phi_1$ ($\sim 8.5 \kpc$). The metallicities of Kshir and GD-1 are compared in Figure~\ref{fig:Fig_chemistry}a. We find ${\rm [Fe/H]}= -1.78\pm0.21$ dex for Kshir, implying that it is systematically more metal-rich than GD-1 by $\sim 0.4$ dex. On performing a two-sample Kolmogorov-Smirnov (KS) test for the null hypothesis that the two [Fe/H] samples are drawn from the same distribution, the resulting probability was found to be $p_{\rm KS}=3.78\times10^{-5}$: indicating that the hypothesis can be rejected at the $4\sigma$ level. Their stellar populations are compared in Figure~\ref{fig:Fig_chemistry}b, where we display the magnitudes corrected for the distance of the stars, as estimated from the orbit models at the corresponding value of $\phi_1$. While the CMDs appear similar, the metallicity distributions suggest that the stellar populations are not identical.

\section{Orbit}\label{sec:orbits}

In MI19 we implemented an orbit-fitting procedure to a sample of GD-1 stars in order to constrain the gravitational potential of the Milky Way. This orbit is shown in Fig~\ref{fig:Fig_1_new}. Here we fix the Galactic potential model derived in that study (which has a circular velocity at the Solar radius of $V_{\rm circ}(R_{\odot}) = 244\kms$, and a density flattening of the dark halo as $q_{\rho}= 0.82$), and follow a similar procedure (with identical likelihood function) as that presented in MI19 to fit the orbit of Kshir. 

\begin{figure}
\begin{center}
\includegraphics[width=\hsize]{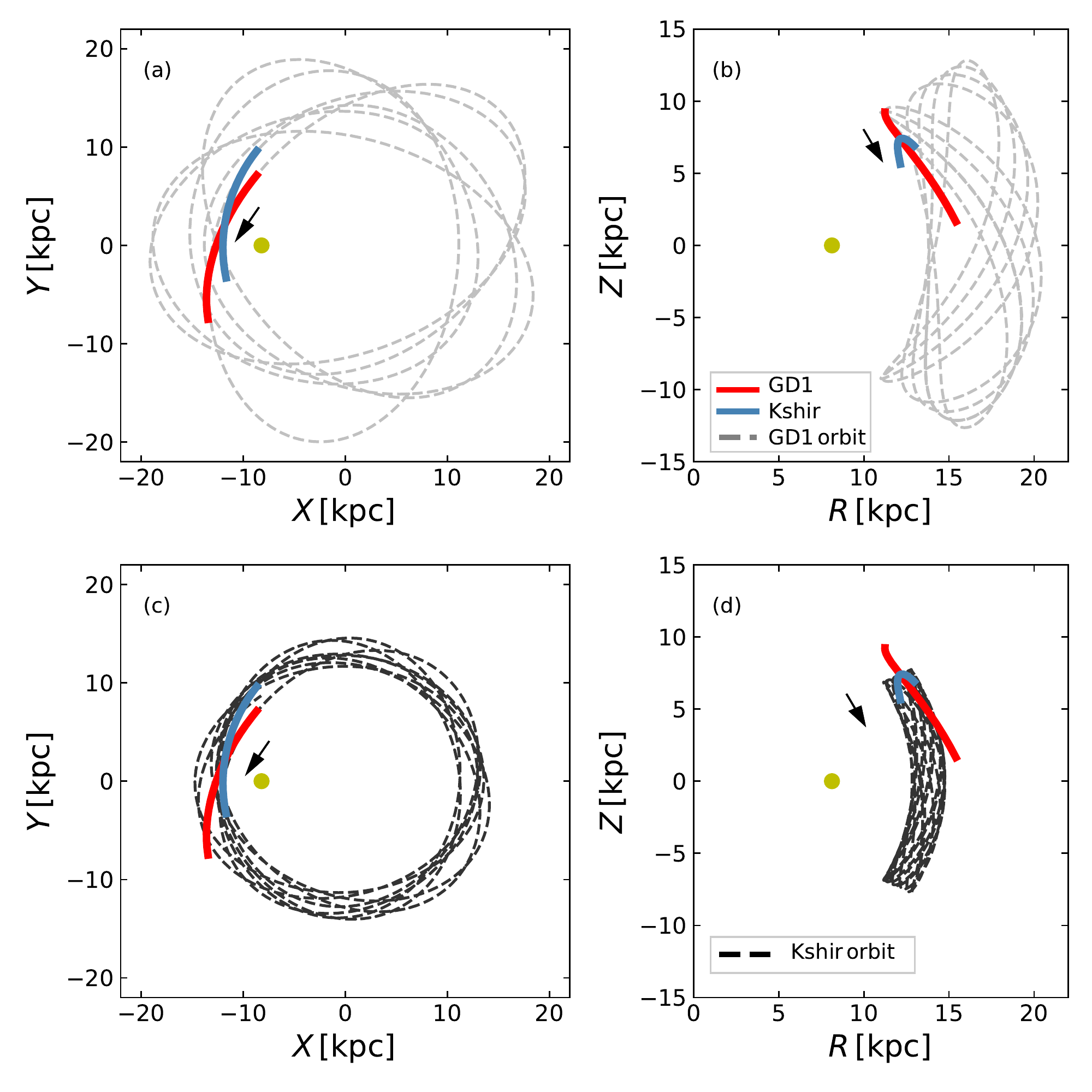}
\end{center}
\vspace{-0.4cm}
\caption{The orbital trajectories of GD-1 and Kshir in the Galactocentric Cartesian system. {\it Top panels:} GD-1's orbit (silver). For perspective, the current locations of GD-1 (red) and Kshir (blue) are also shown. The Galactic centre lies at the origin and the Sun (yellow dot) is at (X,Y,Z)$= (−8.122,0,0) \kpc$. The orbit was integrated backwards in time for 3 Gyr. The arrows represent the direction of motion of the structures. {\it Bottom panels:} As above, but showing Kshir's orbit (black).}
\label{fig:Fig_orbits}
\end{figure}

We used the $42$ Kshir stars identified by the \texttt{STREAMFINDER} (shown as blue points in Fig~\ref{fig:Fig_1_new}), in combination with the $3$ velocity measurements from CFHT and SEGUE that are available for those targets\footnote{We dealt with the missing $v_{\rm los}$ information for the remaining $39$ stars by setting them all to $v_{\rm los}=0\kms$, but with a Gaussian uncertainty of $10^4\kms$. The results are almost identical if instead an uncertainty of $10^3\kms$ is assumed. The choice of adopting a $10^4\kms$ uncertainty is effectively imposing a prior that the stars must be located in the local universe.}. The best-fit orbit for Kshir is shown in Fig~\ref{fig:Fig_1_new}. We find its orbit to be more circular than that of GD-1 (see Fig~\ref{fig:Fig_orbits}), but with $L_z (\sim 2700\pm 200 \kpc\kms)$ comparable to that of GD-1 ($L_z \sim 2950 \kpc\kms$, MI19). The difference in the $L_z$ values stems from the aforementioned offset in the kinematic measurements between the two structures.

These orbits suggest that the last closest approach between Kshir and GD-1 occurred $\sim 1.7\Gyr$ ago with an impact parameter of $\sim 1-2\kpc$, although we caution that these values depend on the assumed Galactic potential model.

\begin{table}
\caption{Spectroscopically confirmed members of Kshir. The sky coordinates are from Gaia DR2, while the $v_{\rm los}$ and [Fe/H] are measurements from CFHT (C), SEGUE (S) and LAMOST (L).}
\vspace{-0.4cm}
\label{tab:GD1x_inventory}
\begin{center}
\begin{tabular}{ccccc}
\hline
RA J2000 & Dec J2000 & $\rm{v_{los}}$ & ${\rm [Fe/H]}$  & Source \\

[deg] & [deg] & $[\kms]$ & $[\rm{dex}]$  & \\
\hline

201.53579  &  67.28841  &  -206.32  &  -1.78 &     S\\
205.87918  &  67.57526  &  -249.95  &    $-$    &     C\\
230.38107  &  68.16672  &  -284.78  &    $-$    &     C\\
\hline
147.96434 	&	9.65499 	&	129.28 	&	-1.56 	&	L\\
150.32399 	&	23.73448 	&	45.09 	&	-1.76 	&	L\\
153.05409 	&	25.1087 	&	43.33 	&	-1.74 	&	S\\
153.10830 	&	26.59341 	&	74.75 	&	-1.41 	&	L\\
154.73641 	&	37.16519 	&	-58.23 	&	-1.47 	&	L\\
156.31086 	&	39.28787 	&	-54.95 	&	-1.94 	&	S\\
156.64035 	&	27.9157 	&	41.96 	&	-1.71 	&	S\\
157.84587 	&	30.59822 	&	-10.42 	&	-2.1 	&	L\\
158.26272 	&	32.5852 	&	-8.47 	&	-1.67 	&	S\\
161.53675 	&	43.45613 	&	-60.52 	&	-2.13 	&	S\\
162.14438 	&	43.73839 	&	-64.03 	&	-1.99 	&	S\\
199.92916 	&	66.09159 	&	-218.86 &	$-$ 	&	S\\
210.43905 	&	66.11245 	&	-253.76 &	-1.83 	&	S\\

\hline
\end{tabular}
\end{center}
\end{table}

%
\section{Discussion and Conclusions}\label{sec:Discussion}

We have presented the discovery of a new stream structure, referred to as ``Kshir'', which criss-crosses the well studied GD-1 stream on the sky, lies at similar distance, and possesses very similar kinematics (Fig~\ref{fig:Fig_1_new}). Moreover, we find Kshir to be also an old and metal poor structure (${\rm [Fe/H] \approx -1.78\pm0.21}$ dex), though slightly more metal rich than GD-1 (Fig~\ref{fig:Fig_chemistry}). Fig~\ref{fig:Fig_1_new}a shows that Kshir's orbit intersects GD-1 at $\phi_1 \sim -20\deg$, which is also the location of the tentative ``gap'' present along GD-1 \citep{WhelanBonacaGD12018}. 

Such a phase-space entanglement in GC streams has not previously been reported. This makes the nature and origin of Kshir, and its possible association with GD-1, very intriguing. We consider below three explanations for the observed correlation.

{\it A) Orbital wraps of same structure.} If Kshir and GD-1 stemmed from the same GC progenitor, then the observed configuration could be due to the presence of different orbital wraps. In this case, Kshir might possibly be a portion of the GD-1 stream that is wrapped by $\sim 360\deg$ (or multiples thereof). The fact that Kshir does not simply line up along the orbit of GD-1 (Fig~\ref{fig:Fig_orbits}) argues against this possibility, although we stress that this result is valid only for the adopted potential derived in MI19. It is possible that other, more complex, Galactic potentials could simultaneously fit extant data and allow Kshir to be a simple wrap of GD-1. Nevertheless, this scenario seems unlikely, due to the difference in the chemical composition of the two structures. This consideration does not completely rule out the scenario, however, since it is possible that the progenitor satellite may have possessed a  radial metallicity gradient (like, e.g., $\omega$ Cen, \citealt{Johnson2010_omegacen}). If tides act slowly, they disrupt the progenitor by progressively removing its outskirts, which can then result in the tidal stream possessing  metallicity variations along its length.

{\it B) A chance alignment of two independent GCs.} It is possible that Kshir and GD-1 are tidal debris of two unassociated GCs. However, given the vast phase-space volume of the Milky Way, such a degree of phase-space overlapping of two unrelated stellar substructures is a low likelihood event. To obtain an estimate of the corresponding probability, we implemented the following test. Employing the \texttt{HaloTools} package \citep{Hearin2017_halotools}, we randomly generated a phase-space distribution in an isotropic NFW halo profile (\citealt{NFW1996}), with a virial mass of $1.28\times 10^{12}\msun$ \citep{Watkins2019}. From this distribution we calculated the probability that two randomly-drawn tracer particles at the distance of the objects of interest have similar orbits. Concretely, we drew $10^4$ random pairs of particles in the Galactocentric distance range between $13$--$15\kpc$ and counted the number of times these pairs possessed a relative difference in the z-component of angular momentum of $\Delta L_z <300 \kpc \kms$, and with relative difference in the energy per unit mass of $\Delta E<10000\,{\rm km^2 s^{-2}}$ (as is the case of Khsir and GD-1). We found this probability to be $\sim 0.007$. This implies that if Kshir and GD-1 are unrelated, then the probability of their chance phase-space alignment is $< 1\%$.

{\it C) Common origin.} The degree of phase-space correlation as we observe for Kshir and GD-1 is possible if the two structures originated from a common parent sub-halo that was accreted onto the Milky Way. Under this scenario, the stellar contents of the sub-halo system would be deposited on very similar orbits. Cosmological simulations show that GCs that evolve within their dark sub-halos, and later accrete into the host halo, give rise to stellar streams that possess substantial morphological complexity \citep{Carlberg2018-StreamSimulation}. This often results in secondary stellar components that accompany the primary GC stream track (see figures in \citealt{Carlberg2018-DensityIn_Stream}), with the overall structure remaining kinematically coherent (much like we see here for Kshir and GD-1). In addition to this, the cosmological simulations further show that the accreted GC streams should lie embedded in broader and dispersed star streams. The reason for this is that the primary stream (which is thin and dense) that survives to the present day is formed once the GC escapes the parent sub-halo and is deposited into the main halo; whereas the wider stellar component enveloping the thin stream is the relic of the stars that were continually removed from the GC while it remained in its parent dark sub-halo. Evidence for this broader stream in GD-1 (labelled ``cocoon'' in Fig~\ref{fig:Fig_1_new}a) was already reported in \citet{MalhanCocoonDetection2019}. Moreover, these simulations also show criss-crossing of streams from separate GCs that formed in a single sub-halo; although an implication is that the sub-halo must be sufficiently massive to host multiple GCs. Therefore, both the Kshir structure and the existence of ``cocoon'' in neighbourhood of GD-1 strengthen the case for this ``accretion'' scenario. Further, the measured difference in metallicity between the two structures argues against a single GC progenitor, although again the constraint is not entirely conclusive (partially because the progenitor could have possessed a radial metallicity gradient). Alternatively, GD-1 and Kshir may correspond to stellar debris produced from either different GC members of the same dark sub-halo, or Kshir may perhaps be debris of the stellar component of the dwarf galaxy that was stripped off during the accretion. A detailed chemical abundance analysis will help to distinguish between these possibilities. Assuming the hypothesized dwarf galaxy hosted a metal poor GC and a field population with ${\rm [Fe/H]_{\rm field}\sim-1.78}$ dex, we estimate its total stellar mass as $M_{*}\sim 10^5 \msun$ (using the Stellar Mass-Stellar Metallicity Relation from \citealt{Kirby2013}), implying $M_{\rm halo}\sim 10^{8-9}\msun$ (from stellar-to-halo-mass relation from \citealt{Read2017}). This makes the progenitor very similar to Eridanus II dwarf \citep{Bechtol2015} which is also known to host a single GC \citep{Crnojevi2016}. However, if the dwarf hosted two GCs (and assuming the field stars to be at least as metal-rich as the GCs), then the above quoted mass values in this case would represent typical lower bounds. None of the luminous satellites have orbits close to GD-1's trajectory \citep{Bonaca_spur_2018}, however, a future detection of a faint disrupting galaxy along GD-1's/ Kshir's orbit would serve as ``smoking gun'' evidence for the proposed accretion scenario. On the other hand, its lack supports a scenario where Kshir and GD-1 were perhaps accreted as GC(s) within an empty dark sub-halo.

The stellar streams of the Milky Way have unexpectedly revealed their rather complex morphologies \citep{WhelanBonacaGD12018, MalhanCocoonDetection2019, Bonaca2019Jhelum}, which presents evidence for a formation mechanism that appears incompatible with a simple tidal disruption model. Modeling GD-1, particularly in light of these new observational constraints, may allow us to develop an understanding of the origins of these recently-found complexities in this well-studied system. This may also allow us to examine whether some GCs can form in otherwise empty CDM sub-halos before, or shortly after re-ionization began (e.g., \citealt{Peebles1984, Mashchenko2005, Ricotti2016}), giving rise to stream structures that exhibit multiple structural components. A detailed dynamical and chemical analysis of GD-1 and Kshir may potentially be useful in distinguishing in-situ and accreted GC stream, and in probing the initial conditions of the dark sub-halo within which they came. \\
 
%
We thank the staff of the CFHT for taking the ESPaDOnS data used here, and for their continued support throughout the project. 

The authors would like to acknowledge the constructive set of comments from the anonymous reviewer. We further thank Monica Valluri and Justin I. Read for helpful conversations. K.M. acknowledges  support  by the $\rm{Vetenskapsr\mathring{a}de}$t (Swedish Research Council) through contract No.  638-2013-8993 and the Oskar Klein Centre for Cosmoparticle Physics, and is grateful for the hospitality received at Observatoire Astronomique de Strasbourg (UdS) where part of the work was performed.

This work has made use of data from the European Space Agency (ESA) mission {\it Gaia} (\url{https://www.cosmos.esa.int/gaia}), processed by the {\it Gaia} Data Processing and Analysis Consortium (DPAC, \url{https://www.cosmos.esa.int/web/gaia/dpac/consortium}). Funding for the DPAC has been provided by national institutions, in particular the institutions participating in the {\it Gaia} Multilateral Agreement. 

Guoshoujing Telescope (the Large Sky Area Multi-Object Fiber Spectroscopic Telescope LAMOST) is a National Major Scientific Project built by the Chinese Academy of Sciences. Funding for the project has been provided by the National Development and Reform Commission. LAMOST is operated and managed by the National Astronomical Observatories, Chinese Academy of Sciences. 

Funding for SDSS-III has been provided by the Alfred P. Sloan Foundation, the Participating Institutions, the National Science Foundation, and the U.S. Department of Energy Office of Science. The SDSS-III web site is \url{http://www.sdss3.org/}.

SDSS-III is managed by the Astrophysical Research Consortium for the Participating Institutions of the SDSS-III Collaboration including the University of Arizona, the Brazilian Participation Group, Brookhaven National Laboratory, Carnegie Mellon University, University of Florida, the French Participation Group, the German Participation Group, Harvard University, the Instituto de Astrofisica de Canarias, the Michigan State/Notre Dame/JINA Participation Group, Johns Hopkins University, Lawrence Berkeley National Laboratory, Max Planck Institute for Astrophysics, Max Planck Institute for Extraterrestrial Physics, New Mexico State University, New York University, Ohio State University, Pennsylvania State University, University of Portsmouth, Princeton University, the Spanish Participation Group, University of Tokyo, University of Utah, Vanderbilt University, University of Virginia, University of Washington, and Yale University.

\bibliographystyle{apj}
\bibliography{ref1}

\begin{thebibliography}{}
\expandafter\ifx\csname natexlab\endcsname\relax\def\natexlab#1{#1}\fi

\bibitem[{{Balbinot} {et~al.}(2016){Balbinot}, {Yanny}, {Li}, {Santiago},
  {Marshall}, {Finley}, {Pieres}, {Abbott}, {Abdalla}, \&
  {Allam}}]{Balbinot2016Phoenix}
{Balbinot}, E., {Yanny}, B., {Li}, T.~S., {et~al.} 2016, \apj, 820, 58

\bibitem[{{Bechtol} {et~al.}(2015){Bechtol}, {Drlica-Wagner}, {Balbinot},
  {Pieres}, {Simon}, {Yanny}, {Santiago}, {Wechsler}, {Frieman}, {Walker},
  {Williams}, {Rozo}, {Rykoff}, {Queiroz}, {Luque}, {Benoit-L{\'e}vy},
  {Tucker}, {Sevilla}, {Gruendl}, {da Costa}, {Fausti Neto}, {Maia}, {Abbott},
  {Allam}, {Armstrong}, {Bauer}, {Bernstein}, {Bernstein}, {Bertin}, {Brooks},
  {Buckley-Geer}, {Burke}, {Carnero Rosell}, {Castander}, {Covarrubias},
  {D'Andrea}, {DePoy}, {Desai}, {Diehl}, {Eifler}, {Estrada}, {Evrard},
  {Fernandez}, {Finley}, {Flaugher}, {Gaztanaga}, {Gerdes}, {Girardi},
  {Gladders}, {Gruen}, {Gutierrez}, {Hao}, {Honscheid}, {Jain}, {James},
  {Kent}, {Kron}, {Kuehn}, {Kuropatkin}, {Lahav}, {Li}, {Lin}, {Makler},
  {March}, {Marshall}, {Martini}, {Merritt}, {Miller}, {Miquel}, {Mohr},
  {Neilsen}, {Nichol}, {Nord}, {Ogando}, {Peoples}, {Petravick}, {Plazas},
  {Romer}, {Roodman}, {Sako}, {Sanchez}, {Scarpine}, {Schubnell}, {Smith},
  {Soares-Santos}, {Sobreira}, {Suchyta}, {Swanson}, {Tarle}, {Thaler},
  {Thomas}, {Wester}, {Zuntz}, \& {DES Collaboration}}]{Bechtol2015}
{Bechtol}, K., {Drlica-Wagner}, A., {Balbinot}, E., {et~al.} 2015, \apj, 807,
  50

\bibitem[{{Belokurov} {et~al.}(2006){Belokurov}, {Zucker}, {Evans}, {Gilmore},
  {Vidrih}, {Bramich}, {Newberg}, {Wyse}, {Irwin}, {Fellhauer}, {Hewett},
  {Walton}, {Wilkinson}, {Cole}, {Yanny}, {Rockosi}, {Beers}, {Bell},
  {Brinkmann}, {Ivezi{\'c}}, \& {Lupton}}]{Belokurov2006}
{Belokurov}, V., {Zucker}, D.~B., {Evans}, N.~W., {et~al.} 2006, \apj, 642,
  L137

\bibitem[{{Bernard} {et~al.}(2016){Bernard}, {Ferguson}, {Schlafly}, {Martin},
  {Rix}, {Bell}, {Finkbeiner}, {Goldman}, {Mart{\'{\i}}nez-Delgado}, {Sesar},
  {Wyse}, {Burgett}, {Chambers}, {Draper}, {Hodapp}, {Kaiser}, {Kudritzki},
  {Magnier}, {Metcalfe}, {Wainscoat}, \& {Waters}}]{Bernard2016}
{Bernard}, E.~J., {Ferguson}, A.~M.~N., {Schlafly}, E.~F., {et~al.} 2016,
  \mnras, 463, 1759

\bibitem[{{Bonaca} {et~al.}(2019{\natexlab{a}}){Bonaca}, {Conroy},
  {Price-Whelan}, \& {Hogg}}]{Bonaca2019Jhelum}
{Bonaca}, A., {Conroy}, C., {Price-Whelan}, A.~M., \& {Hogg}, D.~W.
  2019{\natexlab{a}}, \apjl, 881, L37

\bibitem[{{Bonaca} {et~al.}(2019{\natexlab{b}}){Bonaca}, {Hogg},
  {Price-Whelan}, \& {Conroy}}]{Bonaca_spur_2018}
{Bonaca}, A., {Hogg}, D.~W., {Price-Whelan}, A.~M., \& {Conroy}, C.
  2019{\natexlab{b}}, \apj, 880, 38

\bibitem[{{Bressan} {et~al.}(2012){Bressan}, {Marigo}, {Girardi}, {Salasnich},
  {Dal Cero}, {Rubele}, \& {Nanni}}]{Parsec_isochrones2012}
{Bressan}, A., {Marigo}, P., {Girardi}, L., {et~al.} 2012, \mnras, 427, 127

\bibitem[{{Carlberg}(2018{\natexlab{a}})}]{Carlberg2018-StreamSimulation}
{Carlberg}, R.~G. 2018{\natexlab{a}}, \apj, 861, 69

\bibitem[{{Carlberg}(2018{\natexlab{b}})}]{Carlberg2018-DensityIn_Stream}
---. 2018{\natexlab{b}}, arXiv e-prints, arXiv:1811.10084

\bibitem[{Crnojevi{\'{c}} {et~al.}(2016)Crnojevi{\'{c}}, Sand, Zaritsky,
  Spekkens, Willman, \& Hargis}]{Crnojevi2016}
Crnojevi{\'{c}}, D., Sand, D.~J., Zaritsky, D., {et~al.} 2016, The
  Astrophysical Journal, 824, L14

\bibitem[{Dehnen {et~al.}(2004)Dehnen, Odenkirchen, Grebel, \&
  Rix}]{Dehnen2004}
Dehnen, W., Odenkirchen, M., Grebel, E.~K., \& Rix, H.-W. 2004, The
  Astronomical Journal, 127, 2753

\bibitem[{{Donati} {et~al.}(1997){Donati}, {Semel}, {Carter}, {Rees}, \&
  {Collier Cameron}}]{1997MNRAS.291..658D}
{Donati}, J.-F., {Semel}, M., {Carter}, B.~D., {Rees}, D.~E., \& {Collier
  Cameron}, A. 1997, \mnras, 291, 658

\bibitem[{{Gaia Collaboration} {et~al.}(2018){Gaia Collaboration}, {Brown, A.
  G. A.}, {Vallenari, A.}, {Prusti, T.}, {de Bruijne, J. H. J.}, \& {et
  al.}}]{GaiaDR2_2018_Brown}
{Gaia Collaboration}, {Brown, A. G. A.}, {Vallenari, A.}, {et~al.} 2018, A\&A,
  doi:10.1051/0004-6361/201833051

\bibitem[{{Grillmair}(2009)}]{Grillmair2009_fourStreams}
{Grillmair}, C.~J. 2009, \apj, 693, 1118

\bibitem[{{Grillmair} \& {Carlin}(2016)}]{GrillmairCarlin2016}
{Grillmair}, C.~J., \& {Carlin}, J.~L. 2016, in Astrophysics and Space Science
  Library, Vol. 420, Tidal Streams in the Local Group and Beyond, ed. H.~J.
  {Newberg} \& J.~L. {Carlin}, 87

\bibitem[{{Grillmair} \& {Dionatos}(2006)}]{GrillmairGD12006}
{Grillmair}, C.~J., \& {Dionatos}, O. 2006, \apjl, 643, L17

\bibitem[{{Hearin} {et~al.}(2017){Hearin}, {Campbell}, {Tollerud}, {Behroozi},
  {Diemer}, {Goldbaum}, {Jennings}, {Leauthaud}, {Mao}, {More}, {Parejko},
  {Sinha}, {Sip{\"o}cz}, \& {Zentner}}]{Hearin2017_halotools}
{Hearin}, A.~P., {Campbell}, D., {Tollerud}, E., {et~al.} 2017, \aj, 154, 190

\bibitem[{{Ibata} {et~al.}(2001){Ibata}, {Irwin}, {Lewis}, \&
  {Stolte}}]{Ibata2001Sgr}
{Ibata}, R., {Irwin}, M., {Lewis}, G.~F., \& {Stolte}, A. 2001, \apjl, 547,
  L133

\bibitem[{{Ibata} {et~al.}(2019){Ibata}, {Malhan}, \&
  {Martin}}]{Ibata_Norse_streams2019}
{Ibata}, R.~A., {Malhan}, K., \& {Martin}, N.~F. 2019, \apj, 872, 152

\bibitem[{{Johnson} \& {Pilachowski}(2010)}]{Johnson2010_omegacen}
{Johnson}, C.~I., \& {Pilachowski}, C.~A. 2010, \apj, 722, 1373

\bibitem[{{Kirby} {et~al.}(2013){Kirby}, {Cohen}, {Guhathakurta}, {Cheng},
  {Bullock}, \& {Gallazzi}}]{Kirby2013}
{Kirby}, E.~N., {Cohen}, J.~G., {Guhathakurta}, P., {et~al.} 2013, \apj, 779,
  102

\bibitem[{{Koposov} {et~al.}(2010){Koposov}, {Rix}, \& {Hogg}}]{Koposov2010}
{Koposov}, S.~E., {Rix}, H.-W., \& {Hogg}, D.~W. 2010, \apj, 712, 260

\bibitem[{{Kravtsov} \& {Gnedin}(2005)}]{Kravtsov2005}
{Kravtsov}, A.~V., \& {Gnedin}, O.~Y. 2005, \apj, 623, 650

\bibitem[{{Kruijssen}(2014)}]{Kruijssen2014}
{Kruijssen}, J.~M.~D. 2014, Classical and Quantum Gravity, 31, 244006

\bibitem[{{Lindegren, L.} {et~al.}(2018){Lindegren, L.}, {Hernandez, J.},
  {Bombrun, A.}, {Klioner, S.}, {Bastian, U.}, \& {Ramos-Lerate,
  M.}}]{GaiaDR2_2018_astrometry}
{Lindegren, L.}, {Hernandez, J.}, {Bombrun, A.}, {et~al.} 2018, A\&A,
  doi:10.1051/0004-6361/201832727

\bibitem[{{Malhan} \& {Ibata}(2018)}]{Malhan2018_SF}
{Malhan}, K., \& {Ibata}, R.~A. 2018, \mnras, 477, 4063

\bibitem[{{Malhan} \& {Ibata}(2019)}]{Malhan2018PotentialGD1}
---. 2019, \mnras, 486, 2995

\bibitem[{{Malhan} {et~al.}(2019){Malhan}, {Ibata}, {Carlberg}, {Valluri}, \&
  {Freese}}]{MalhanCocoonDetection2019}
{Malhan}, K., {Ibata}, R.~A., {Carlberg}, R.~G., {Valluri}, M., \& {Freese}, K.
  2019, The Astrophysical Journal, 881, 106

\bibitem[{{Malhan} {et~al.}(2018){Malhan}, {Ibata}, \&
  {Martin}}]{Malhan_Ghostly_2018}
{Malhan}, K., {Ibata}, R.~A., \& {Martin}, N.~F. 2018, \mnras, 481, 3442

\bibitem[{{Mashchenko} \& {Sills}(2005)}]{Mashchenko2005}
{Mashchenko}, S., \& {Sills}, A. 2005, \apj, 619, 243

\bibitem[{{Myeong} {et~al.}(2017){Myeong}, {Jerjen}, {Mackey}, \& {Da
  Costa}}]{Myeong2017_Streams}
{Myeong}, G.~C., {Jerjen}, H., {Mackey}, D., \& {Da Costa}, G.~S. 2017, \apjl,
  840, L25

\bibitem[{{Navarro} {et~al.}(1996){Navarro}, {Frenk}, \& {White}}]{NFW1996}
{Navarro}, J.~F., {Frenk}, C.~S., \& {White}, S.~D.~M. 1996, \apj, 462, 563

\bibitem[{{Peebles}(1984)}]{Peebles1984}
{Peebles}, P.~J.~E. 1984, \apj, 277, 470

\bibitem[{{Price-Whelan} \& {Bonaca}(2018)}]{WhelanBonacaGD12018}
{Price-Whelan}, A.~M., \& {Bonaca}, A. 2018, \apjl, 863, L20

\bibitem[{{Read} {et~al.}(2017){Read}, {Iorio}, {Agertz}, \&
  {Fraternali}}]{Read2017}
{Read}, J.~I., {Iorio}, G., {Agertz}, O., \& {Fraternali}, F. 2017, \mnras,
  467, 2019

\bibitem[{{Renaud} {et~al.}(2017){Renaud}, {Agertz}, \& {Gieles}}]{Renaud2017}
{Renaud}, F., {Agertz}, O., \& {Gieles}, M. 2017, \mnras, 465, 3622

\bibitem[{{Ricotti} {et~al.}(2016){Ricotti}, {Parry}, \&
  {Gnedin}}]{Ricotti2016}
{Ricotti}, M., {Parry}, O.~H., \& {Gnedin}, N.~Y. 2016, \apj, 831, 204

\bibitem[{{Robin} {et~al.}(2012){Robin}, {Luri}, {Reyl{\'e}}, {Isasi}, {Grux},
  {Blanco-Cuaresma}, {Arenou}, {Babusiaux}, {Belcheva}, {Drimmel}, {Jordi},
  {Krone-Martins}, {Masana}, {Mauduit}, {Mignard}, {Mowlavi},
  {Rocca-Volmerange}, {Sartoretti}, {Slezak}, \& {Sozzetti}}]{GUMS2012_Robin}
{Robin}, A.~C., {Luri}, X., {Reyl{\'e}}, C., {et~al.} 2012, \aap, 543, A100

\bibitem[{{Schlegel} {et~al.}(1998){Schlegel}, {Finkbeiner}, \&
  {Davis}}]{Schlegel1998}
{Schlegel}, D.~J., {Finkbeiner}, D.~P., \& {Davis}, M. 1998, \apj, 500, 525

\bibitem[{{Shipp} {et~al.}(2018){Shipp}, {Drlica-Wagner}, {Balbinot},
  {Ferguson}, {Erkal}, {Li}, {Bechtol}, {Belokurov}, {Buncher}, {Carollo},
  {Carrasco Kind}, {Kuehn}, {Marshall}, {Pace}, {Rykoff}, {Sevilla-Noarbe},
  {Sheldon}, {Strigari}, {Vivas}, {Yanny}, {Zenteno}, {Abbott}, {Abdalla},
  {Allam}, {Avila}, {Bertin}, {Brooks}, {Burke}, {Carretero}, {Castander},
  {Cawthon}, {Crocce}, {Cunha}, {D'Andrea}, {da Costa}, {Davis}, {De Vicente},
  {Desai}, {Diehl}, {Doel}, {Evrard}, {Flaugher}, {Fosalba}, {Frieman},
  {Garc{\'\i}a-Bellido}, {Gaztanaga}, {Gerdes}, {Gruen}, {Gruendl}, {Gschwend},
  {Gutierrez}, {Hartley}, {Honscheid}, {Hoyle}, {James}, {Johnson}, {Krause},
  {Kuropatkin}, {Lahav}, {Lin}, {Maia}, {March}, {Martini}, {Menanteau},
  {Miller}, {Miquel}, {Nichol}, {Plazas}, {Romer}, {Sako}, {Sanchez},
  {Santiago}, {Scarpine}, {Schindler}, {Schubnell}, {Smith}, {Smith},
  {Sobreira}, {Suchyta}, {Swanson}, {Tarle}, {Thomas}, {Tucker}, {Walker},
  {Wechsler}, \& {DES Collaboration}}]{Shipp2018}
{Shipp}, N., {Drlica-Wagner}, A., {Balbinot}, E., {et~al.} 2018, \apj, 862, 114

\bibitem[{{Watkins} {et~al.}(2019){Watkins}, {van der Marel}, {Sohn}, \&
  {Evans}}]{Watkins2019}
{Watkins}, L.~L., {van der Marel}, R.~P., {Sohn}, S.~T., \& {Evans}, N.~W.
  2019, \apj, 873, 118

\bibitem[{{Yanny} {et~al.}(2009){Yanny}, {Rockosi}, {Newberg}, {Knapp},
  {Adelman-McCarthy}, {Alcorn}, {Allam}, {Allende Prieto}, {An}, {Anderson},
  {Anderson}, {Bailer-Jones}, {Bastian}, {Beers}, {Bell}, {Belokurov},
  {Bizyaev}, {Blythe}, {Bochanski}, {Boroski}, {Brinchmann}, {Brinkmann},
  {Brewington}, {Carey}, {Cudworth}, {Evans}, {Evans}, {Gates}, {G{\"a}nsicke},
  {Gillespie}, {Gilmore}, {Nebot Gomez-Moran}, {Grebel}, {Greenwell}, {Gunn},
  {Jordan}, {Jordan}, {Harding}, {Harris}, {Hendry}, {Holder}, {Ivans},
  {Ivezi{\v c}}, {Jester}, {Johnson}, {Kent}, {Kleinman}, {Kniazev},
  {Krzesinski}, {Kron}, {Kuropatkin}, {Lebedeva}, {Lee}, {French Leger},
  {L{\'e}pine}, {Levine}, {Lin}, {Long}, {Loomis}, {Lupton}, {Malanushenko},
  {Malanushenko}, {Margon}, {Martinez-Delgado}, {McGehee}, {Monet}, {Morrison},
  {Munn}, {Neilsen}, {Nitta}, {Norris}, {Oravetz}, {Owen}, {Padmanabhan},
  {Pan}, {Peterson}, {Pier}, {Platson}, {Re Fiorentin}, {Richards}, {Rix},
  {Schlegel}, {Schneider}, {Schreiber}, {Schwope}, {Sibley}, {Simmons},
  {Snedden}, {Allyn Smith}, {Stark}, {Stauffer}, {Steinmetz}, {Stoughton},
  {SubbaRao}, {Szalay}, {Szkody}, {Thakar}, {Sivarani}, {Tucker}, {Uomoto},
  {Vanden Berk}, {Vidrih}, {Wadadekar}, {Watters}, {Wilhelm}, {Wyse}, {Yarger},
  \& {Zucker}}]{SEGUE_SDSS2009}
{Yanny}, B., {Rockosi}, C., {Newberg}, H.~J., {et~al.} 2009, \aj, 137, 4377

\bibitem[{{Zhao} {et~al.}(2012){Zhao}, {Zhao}, {Chu}, {Jing}, \&
  {Deng}}]{Lamost2012}
{Zhao}, G., {Zhao}, Y., {Chu}, Y., {Jing}, Y., \& {Deng}, L. 2012, ArXiv
  e-prints, arXiv:1206.3569

\end{thebibliography}

\end{document}